\documentclass[12pt,preprint]{aastex}

\usepackage{amsmath}

\slugcomment{}

\shorttitle{About the possible role of hydrocarbon lakes in the origin of Titan's noble gas atmospheric depletion}

\shortauthors{Cordier et al.}

\begin{document}

\title{About the possible role of hydrocarbon lakes in the origin of Titan's noble gas atmospheric depletion}

\author{
D.~Cordier\altaffilmark{1,2,3},
O.~Mousis\altaffilmark{4},
J. I.~Lunine\altaffilmark{5},
S.~Lebonnois\altaffilmark{6},
P.~Lavvas,\altaffilmark{7}
L.Q.~Lobo\altaffilmark{8},
A.G.M.~Ferreira\altaffilmark{8}}

\email{daniel.cordier@ensc-rennes.fr}

\altaffiltext{1}{Ecole Nationale Sup{\'e}rieure de Chimie de Rennes, CNRS, UMR 6226, Avenue du G\'en\' eral Leclerc,
                 CS 50837, 35708 Rennes Cedex 7, France}
                 
\altaffiltext{2}{Universit\'e europ\'eenne de Bretagne, Rennes, France}
                 
\altaffiltext{3}{Institut de Physique de Rennes, CNRS, UMR 6251, Universit{\'e} de Rennes 1, Campus de Beaulieu, 35042 Rennes, France}

\altaffiltext{4}{Universit{\'e} de Franche-Comt{\'e}, Institut UTINAM, CNRS/INSU, UMR 6213, 25030 Besan\c{c}on Cedex, France}

\altaffiltext{5}{Dipartimento di Fisica, Universit{\`a} degli Studi di Roma ``Tor Vergata'', Rome, Italy}

\altaffiltext{6}{Laboratoire de M{\'e}t{\'e}orologie Dynamique, Jussieu, Box 99, 75252 PARIS cedex 05, France}

\altaffiltext{7}{Lunar and Planetary Laboratory, University of Arizona, Tucson, AZ, USA}

\altaffiltext{8}{Departamento de Engenharia Quimica, Universidade de Coimbra, Coimbra 3030-290, Portugal}

\begin{abstract}
{An unexpected feature of Titan's atmosphere is the strong depletion in primordial noble gases revealed by the Gas Chromatograph Mass Spectrometer aboard the \textit{Huygens} probe during its descent on 2005 January 14. Although several plausible explanations have already been formulated, no definitive response to this issue has been yet found. Here, we investigate the possible sequestration of these noble gases in the liquid contained in lakes and wet terrains on Titan and the consequences for their atmospheric abundances. Considering the atmosphere and the liquid existing on the soil as a whole system, we compute the abundance of each noble gas relative to nitrogen. To do so, we make the assumption of thermodynamic equilibrium between the liquid and the atmosphere, the abundances of the different constituents being determined via {\bf } regular solution theory. We find that xenon's atmospheric depletion can be explained by its dissolution at ambient temperature in the liquid presumably present on Titan's soil. In the cases of argon and krypton, we find that the fractions incorporated in the liquid are negligible, implying that an alternative mechanism must be invoked to explain their atmospheric depletion.}
\end{abstract}


\section{\label{intro}Introduction}

A striking feature of the atmosphere of Titan is that no primordial noble gases other than argon were detected by the Gas Chromatograph Mass Spectrometer (GCMS) aboard the \textit{Huygens} probe during its descent to Titan's surface in January 2005. The detected argon includes primordial $^{36}$Ar present in subsolar abundance in Titan's atmosphere ($^{36}$Ar/$^{14}$N is found to be about six orders of magnitude lower than the solar value) and the radiogenic isotope $^{40}$Ar, which is a decay product of $^{40}$K \citep{niemann_etal_2005}. The other primordial noble gases $^{38}$Ar, Kr and Xe were not detected by the GCMS instrument, yielding upper limits of 10$^{-8}$ for their atmospheric mole fractions.

 The interpretation of the noble gas deficiency measured in Titan's atmosphere has been the subject of several studies in the recent literature. Thus, \cite{osegovic_max_2005} proposed that these species could be preferentially stored in clathrates present on the satellite's surface. They calculated the composition of clathrates on the surface of Titan using the program CSMHYD developed by \cite{sloan_1998} and showed that such crystalline ice structures may act as a sink for Xe. However, the CSMHYD code used by \cite{osegovic_max_2005} is not suitable below 140K for gas mixtures of interest whereas the mean surface temperature of Titan is below 95 K \citep{cordier_etal_2009}, and the authors did not explicitly calculate the trapping efficiencies of Ar and Kr in clathrates on the surface of the satellite. These considerations led \cite{thomas_etal_2007,thomas_etal_2008} to rethink their results. In both studies, the authors found that the trapping efficiency of clathrates is high enough to significantly decrease the atmospheric concentrations of Xe and Kr irrespective of the initial gas phase composition, provided that these clathrates are abundant enough on the surface of Titan. In contrast, they found that Ar is poorly trapped in clathrates and that this mechanism alone could not explain the argon impoverishment measured in Titan's atmosphere. Another interpretation of the Ar, Kr and Xe deficiencies is that the haze present in Titan's atmosphere could simultaneously trap these three noble gases in a way consistent with the observed atmospheric abundances \citep{jacovi_bar-nun_2008}. In this mechanism, the open structure of the small aerosol particles would allow the noble gas atoms to fill their pores. All these hypotheses are based on different assumptions (requirement of large amounts of clathrates on the satellite's surface or formation of Titan's aerosols in exactly the same conditions as those used during laboratory experiments) that will need to be investigated by in situ measurements or observations performed by future spacecraft missions.
 
 In this paper, we offer another hypothesis: that Titan's hydrocarbon lakes play a key role in the impoverishment of its atmospheric noble gases. Indeed, hundreds of radar dark features interpreted as hydrocarbon lakes have been detected in the polar regions \citep{stofan_etal_2007}. Recently, \cite{cordier_etal_2009} have published a study of the chemical composition of Titan's lakes, which is based on the direct abundance measurements from the Gas Chromatograph Mass Spectrometer aboard the \textit{Huygens} probe and recent photochemical models based on the vertical temperature profile derived by the \textit{Huygens} Atmospheric Structure Instrument. Here, we extend the model of \cite{cordier_etal_2009} by including simultaneously Ar, Kr and Xe in the composition of the liquid phase. We then explore the amount of liquid that is needed on the surface of Titan to account for the measured noble gas atmospheric abundances assuming solar abundances in the bulk system.

\section{Noble gases sequestration in Titan's surface liquid phase}
\label{sec:method}

Here we consider the atmosphere and the liquid existing on the surface of Titan as a single, fully--coupled system. Our approach consists in i) computing the ratio 
$N_{\rm NG}/N_{\rm nitrogen}$ where $N_{\rm NG}$ and $N_{\rm nitrogen}$ are the total numbers of atoms of a given noble gas NG and of nitrogen, respectively, and ii) comparing this result to the ratio $(N_{\rm NG}/N_{\rm nitrogen})_{\odot}$ derived from protosolar abundances (Lodders 2003). Our calculations always refer to nitrogen because it is the most abundant compound detected in Titan's atmosphere and because thermodynamic equilibrium models predict that it is also present in the lakes 
\citep[see][]{cordier_etal_2009}. Under these conditions, for a noble gas NG, we can write

\begin{equation}\label{firstequation}
  \frac{N_{\rm NG}}{N_{\rm nitrogen}} = \frac{N_{\rm tot,NG}^{\rm(liq)}+N_{\rm tot,NG}^{\rm(atm)}}{N_{\rm tot, nitrogen}^{\rm(liq)}
                                   + N_{\rm tot, nitrogen}^{\rm(atm)}}
\end{equation} 

\noindent where $N_{\rm tot}^{\rm(liq)}$ and $N_{\rm tot}^{\rm(atm)}$ are the total number of atoms of element NG or nitrogen in the liquid and in the atmosphere of Titan, respectively.

The total mole number of element NG in Titan's liquid is given by

\begin{equation}\label{Nliq}
   N_{\rm tot,NG}^{\rm(liq)} = \frac{x_{\rm NG}^{\rm(liq)} \times V_{tot}^{\rm(liq)} \times \rho^{\rm(liq)}}{\bar{M}},
\end{equation}

\noindent  where $x_{\rm NG}^{\rm(liq)}$ is the mole fraction of atoms of NG in lakes computed following the method described in \cite{cordier_etal_2009}, $V_{\rm tot}^{\rm(liq)}$ is the total volume of Titan's liquid in contact with the atmosphere, $\rho^{\rm(liq)}$ their mean density (in kg.m$^{-3}$) and $\bar{M}$ the mean molecular weight of the liquid. $\bar{M}$ is given by

\begin{equation}\label{meanmol}
  \bar{M} = \sum_{j} x_{j} \times M_{j},
\end{equation}

\noindent where the sum $\sum_{j}$ runs over all species present in the liquid phase. The composition of the liquid is calculated via the thermodynamic equilibrium model of Titan's lakes described by \cite{cordier_etal_2009}. In this approach, {\bf} thermodynamic equilibrium, which translates into the equality of chemical potentials for each species listed in Table~\ref{liqabund} from N$_{2}$ to C$_{2}$H$_{6}$ (except{\bf} for
H$_{2}$), can be expressed as

\begin{equation}\label{equa1}
  y_{i} \, P = \Gamma_{i} \, x_{i} \, P_{vp,i},
\end{equation}

\noindent where $P$ is the total pressure at Titan's surface, $y_{i}$ and $x_{i}$ are respectively the mole fractions of the $i$ compound in the atmosphere and the liquid, $P_{vp,i}$ is its vapor pressure, and $\Gamma_{i}$ is its activity coefficient in the liquid determined with Equation 2 of \cite{dubouloz_etal_1989}. Abundances of compounds below C$_2$H$_6$ in Table~\ref{liqabund} are expressed proportionally to that of C$_2$H$_6$ both in the precipitation and in the liquid existing in the lakes or in the putative porous network.

\begin{table}[h]
\caption[]{Composition of liquid (mole fraction at given temperature).}
\begin{center}
\begin{tabular}{lccccccc}
\hline
\hline
            			&  	87 K      			    		&  	90 K       			    			& 93.65 K                		&    		\\
\hline														
 N$_{2}$    		&     	$1.22\times 10^{-2}$			&      $4.94\times 10^{-3}$ 	    		& $2.96\times 10^{-3}$      &    		\\
 CH$_{4}$  		&     	$2.18\times 10^{-1}$ 		&      $9.74\times 10^{-2}$ 	    		& $5.56\times 10^{-2}$      &    		\\
 Ar         			&     	$1.01\times 10^{-5}$ 		&      $4.95\times 10^{-6}$ 	    		& $3.09\times 10^{-6}$      &    		\\
 Xe         			&     	$8.55\times 10^{-3}$ 		&      $1.52\times 10^{-3}$ 	    		& $3.09\times 10^{-4}$      &    		\\
 Kr         			&     	$7.72\times 10^{-9}$ 		&      $3.13\times 10^{-9}$ 	    		& $1.92\times 10^{-9}$      &    		\\
 CO         			&     	$1.24\times 10^{-6}$ 		&      $4.25\times 10^{-7}$ 	    		& $2.05\times 10^{-7}$      &    		\\
 H$_{2}$         		&   $2.94\times 10^{-11}$       		&      $4.08\times 10^{-11}$                   	& $5.12\times 10^{-11}$    &	          \\ 
C$_{2}$H$_{6}$ 	&  	$6.55\times 10^{-1}$ 		&      $7.62\times 10^{-1}$ 	    		& $7.95\times 10^{-1}$      &    		\\
C$_{3}$H$_{8}$ 	&  	$6.36\times 10^{-2}$ 		&      $7.40\times 10^{-2}$ 	   		& $7.71\times 10^{-2}$      &    		\\	
C$_{4}$H$_{8}$ 	&  	$1.19\times 10^{-2}$ 		&      $1.39\times 10^{-2}$ 	    		& $1.45\times 10^{-2}$      &    		\\				
HCN           		&  	$9.06\times 10^{-3}$ 		&      $2.08\times 10^{-2}$ 	  		& $2.89\times 10^{-2}$      &  (s) 	\\
C$_{4}$H$_{10}$ 	& 	$1.04\times 10^{-2}$ 		&      $1.21\times 10^{-2}$ 	 		& $1.26\times 10^{-2}$      & (ns) 	\\
C$_{2}$H$_{2}$  	& 	$9.83\times 10^{-3}$ 		&      $1.14\times 10^{-2}$ 	 		& $1.19\times 10^{-2}$      & (ns) 	\\  
CH$_{3}$CN      	& 	$8.48\times 10^{-4}$ 		&      $9.87\times 10^{-4}$ 	 		& $1.03\times 10^{-3}$      & (ns) 	\\
CO$_{2}$        		& 	$2.50\times 10^{-4}$ 		&      $2.92\times 10^{-4}$ 	 		& $3.04\times 10^{-4}$      & (ns) 	\\
C$_{6}$H$_{6}$  	& 	$1.93\times 10^{-4}$ 		&      $2.24\times 10^{-4}$ 	 		& $2.34\times 10^{-4}$      & (ns) 	\\
     \hline
\end{tabular}
\tablecomments{From HCN to C$_{6}$H$_{6}$, compounds are in the solid state in precipitates and are assumed to dissolve when they reach the liquid phase. (s): saturated; (ns): non saturated. Ar is the total argon contained in all isotopes.}
\end{center}
\label{liqabund}
\end{table} 

In order to compute $N_{\rm tot,NG}^{\rm(liq)}$ for all noble gases, we have extended the thermodynamic equilibrium model by adding Kr and Xe to the list of species already taken into account by \cite{cordier_etal_2009}. The thermodynamic data essentially derive from the NIST database\footnote{\url{http://webbook.nist.gov/chemistry}} in which the vapor pressures are expressed in the forms of Antoine's equations. On the other hand, the enthalpy of vaporization and molar volume of Ar, which are needed in the computation of its activity coefficient via the determination of its solubility parameter \citep[see][]{prausnitz_1986}, have been updated relative to the value adopted by \cite{cordier_etal_2009} and now are derived from the laboratory measurements published by \cite{ferreira_lobo_2008} and \cite{tegeler_etal_1999}, respectively. On the other hand, the enthalpies of vaporization and molar volumes of Kr and Xe all derive from the experimental data published by \cite{ferreira_lobo_2009}.

The total mole number of element NG in Titan's atmosphere is determined via the following vertical integration

\begin{equation}\label{yatm}
    N_{\rm tot,NG}^{\rm(atm)} = \int_{z=0}^{\rm z=H} 
                                \frac{y_{\rm NG}^{\rm(atm)} \times \rho^{\rm(atm)}(z)}{\bar{M}^{\rm(atm)}(z)} 
                                \, 4\pi \, (R_{\rm Titan}+z)^{2}
                                \mathrm{d}z
\end{equation}

\noindent where $R_{\rm Titan}$ is the radius of Titan, $z$ the altitude and $H$ the maximum elevation at which the \textit{Huygens} GCMS started to collect
data, $\rho^{\rm(atm)}(z)$ is the atmospheric density at the elevation $z$ {\bf } whose determination derives from the \textit{Huygens} HASI data \citep{fulchignoni_etal_2005}, $x_{\rm NG}^{\rm(atm)}$ is the mole fraction of element NG derived from the GCMS data at the ground level \citep{niemann_etal_2005} and {\bf is}assumed to be constant whatever the altitude. The mean molecular weight $\bar{M}^{\rm(atm)}(z)$ of the atmosphere is also derived from the \textit{Huygens} GCMS data using an approach similar to the one used to calculate the mean molecular weight of the liquid (see. Eq.~\ref{meanmol}).
\begin{table}[h]
\label{atmoscompo}
\caption[]{Assumed composition of Titan's atmosphere at the ground level.}
\begin{center}
\begin{tabular}{lcc}
\hline
\hline
\noalign{\smallskip}
Atmosphere			& Mole fraction			    		& Determination         			\\
\hline
H$_2$				& $9.8 \times 10^{-4}$			& \textit{Huygens} GCMS$^{(a)}$	\\
N$_2$				& 0.95                          			& This work					\\
CH$_4$				& 0.0492 						& \textit{Huygens} GCMS$^{(b)}$	\\
CO					& $4.70 \times 10^{-5}$			& \textit{Cassini} CIRS$^{(c)}$  		\\
$^{40}$Ar		    		& $4.32 \times 10^{-5}$ 	    		& \textit{Huygens} GCMS$^{(b)}$	\\
$^{36}$Ar		    		& $2.80 \times 10^{-7}$ 	    		& \textit{Huygens} GCMS$^{(b)}$	\\
Kr			    		& $10^{-8}$ 	    				& \textit{Huygens} GCMS$^{(b)}$	\\
Xe				    	& $10^{-8}$ 	    				& \textit{Huygens} GCMS$^{(b)}$	\\
C$_2$H$_6$          		& $1.49 \times 10^{-5}$         		& This work					\\
\hline
\end{tabular}
\tablecomments{$^{(a)}$\cite{owen_niemann_2009}; $^{(b)}$\cite{niemann_etal_2005}; $^{(c)}$\cite{dekok_etal_2007}. N$_2$ and C$_2$H$_6$ abundances are determined from our model (see text).}
\end{center}
\end{table}
\section{Assumptions about the Titan environment}
\label{sec:results}

Atmospheric mole fraction data used as inputs in our model are gathered in Table~\ref{atmoscompo}. The atmospheric abundances of species which are not included in this table are assumed negligible. We have set the atmospheric mole fractions of Kr and Xe to $10^{-8}$ because they correspond to the detection limit of the \textit{Huygens} GCMS instrument. Following the approach of \cite{cordier_etal_2009}, atmospheric mole fractions of N$_{2}$ and C$_{2}$H$_{6}$ are treated as unknowns of our problem and atmospheric abundances of compounds below the one of C$_{2}$H$_{6}$ are neglected. In order to investigate the temperature dependence of the mole fractions of noble gases trapped in Titan's surface liquid, we have allowed the ground temperature to range between 87 and 94 K. These extreme values bracket the mean ground temperature estimated to be $\sim$90--91 K in northern polar regions where lake candidates are located \citep{janssen_etal_2009} and the temperature of 93.65 K measured by the \textit{Huygens} probe at its landing site \citep{niemann_etal_2005}.

Current inventories of Titan's lakes are estimated to range between $3 \times 10^{4}$ and $3 \times 10^{5}$ km$^{3}$ \citep{lorenz_etal_2008}. In the present study, we consider a more generous range for the total volume of liquid existing at the surface of Titan, including a fraction that might reside in wet terrains. A substantial fraction of liquid incorporated in Titan's soil, even at the equator, is plausible because the \textit{Huygens} probe firmly identified ethane in the mass spectra taken from the surface \citep{niemann_etal_2005}. Moreover, it has been proposed that the subsurface of Titan is porous, implying that large amounts of liquid hydrocarbons could remain in contact with the atmosphere via open pores in the soil \citep{mousis_schmitt_2008}. Hence, assuming that the total volume of liquid existing on Titan's surface could be up to 10 times the one estimated for the lakes, we consider here three different liquid volume values: $3 \times 10^{4}$, $3 \times 10^{5}$ and $3 \times 10^{6}$ km$^{3}$. Note that the larger volume of liquid considered here remains $\sim$10 times lower than the total amount of ethane that could have precipitated on Titan's surface if one assumes a precipitation rate of $3.4 \times 10^{9}$ molecules.cm$^{-2}$.s$^{-1}$ \citep{lavvas_etal_2008a,lavvas_etal_2008b} over 4.5 gigayears.

\begin{center}
\rotatebox{-90}{\resizebox{12 cm}{15.5294117647059 cm}{\includegraphics{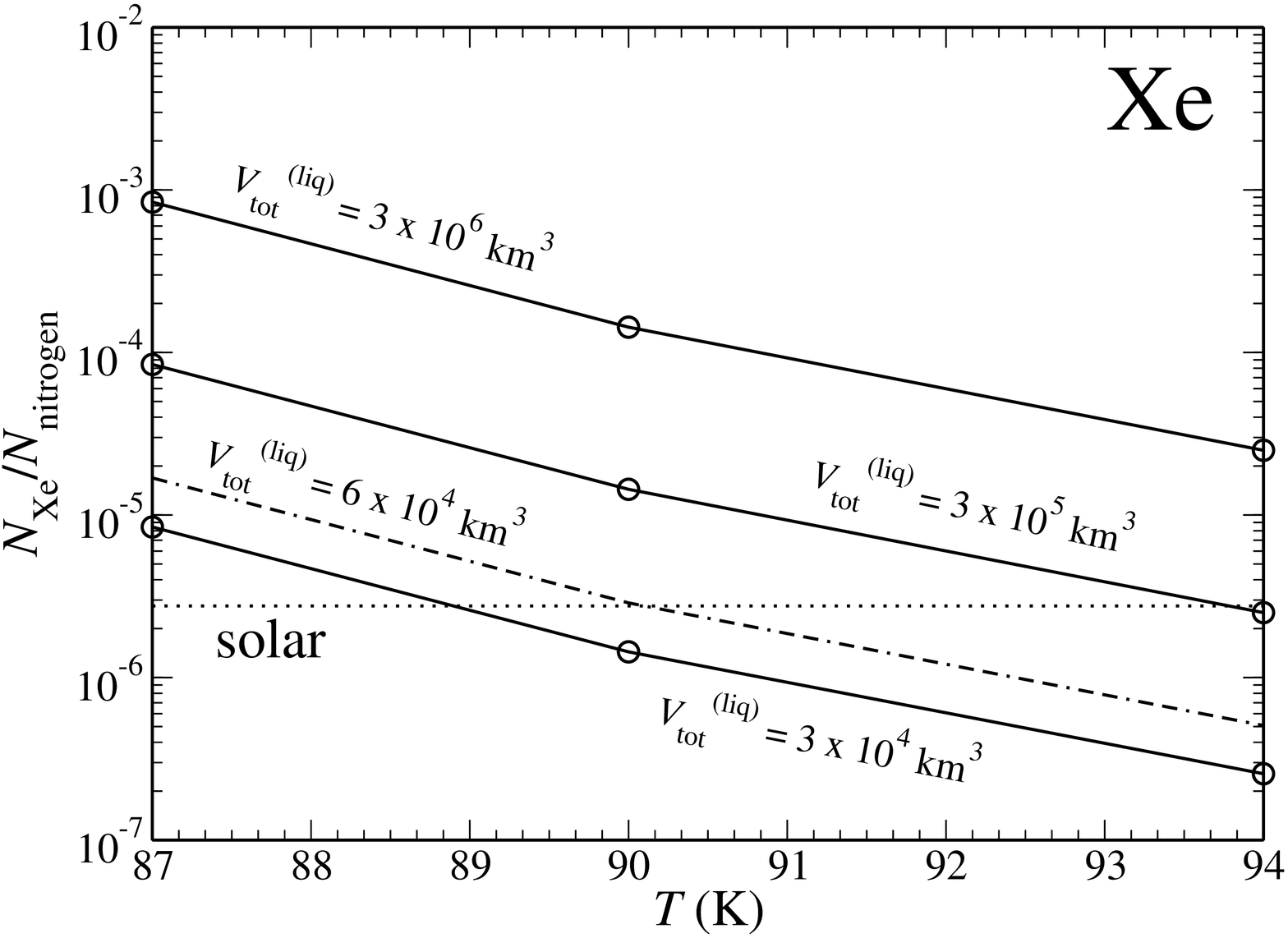}}}

\begin{figure}[h]
\caption{The global $N_{\rm Xe}/N_{\rm nitrogen}$ ratio in Titan versus ground temperature for different volumes of lakes. The horizontal line represents the solar value. Other lines, from top to bottom, derive from our calculations for the considered volumes of liquid available at ground level. The assumed atmospheric mole fraction corresponds to the GCMS detection threshold ($y_{\rm Xe} = 10^{-8}$).}
\label{NsN}
\end{figure}
\end{center}

\section{Results}

Our calculations, including both $^{36}$Ar and $^{40}$Ar isotopes, produce a $N_{\rm Ar}/N_{\rm nitrogen}$ 
ratio ($\sim$$2.24 \times 10^{-5}$) remains more than three orders of magnitude lower than the value inferred from solar abundances ($(N_{\rm Ar}/N_{\rm nitrogen})_{\odot}= 5.26 \times 10^{-2}$), irrespective of the adopted values for the temperature and total volume of lakes. Because $^{36}$Ar is the main primordial argon isotope \citep{lodders_2003}, we note that even if all the argon dissolved in the liquid was in the form of this isotope, this could not explain its apparent depletion in Titan's atmosphere. This behavior is explained by the very low solubility of argon in the C$_2$H$_6$-dominated liquid in the considered temperature range. Even if the liquid volume is of the order of $3 \times 10^{6}$ km$^{3}$, almost all argon remains in the atmosphere, the fraction incorporated in liquid playing only a negligible role. Hence, whatever the source of argon, i.e. primordial or radiogenic, we conclude that the argon depletion observed in Titan's atmosphere cannot be explained by its sequestration in the liquid in contact with Titan's atmosphere.

  In the case of krypton, the computed $N_{\rm Kr}/N_{\rm nitrogen}$ ratio ($\sim$$5 \times 10^{-9}$) is about 3--4 orders of magnitude lower than the solar value ($(N_{\rm Kr}/N_{\rm nitrogen})_{\odot}= 2.83 \times 10^{-5}$), irrespective of the adopted values for the temperature and total volume of lakes. Similarly to the case of argon, dissolution of krypton in liquid hydrocarbon on the surface of Titan cannot explain the observed depletion.

  Figure \ref{NsN} summarizes the behavior of the global ratio $N_{\rm Xe}/N_{\rm nitrogen}$ as a function of the temperature of Titan's surface and of the available volume of liquid. Because the thermodynamic equilibrium between the liquid phase and the atmosphere requires a larger mole fraction of xenon in the liquid, the resulting $N_{\rm Xe}/N_{\rm nitrogen}$ ratio is larger than the one calculated from the detection limit of the GCMS instrument. Moreover, since the atmospheric mole fraction of xenon remains fixed in our system, its solubility diminishes in the liquid with an increase of temperature.

Figure~\ref{NsN} shows that only the $N_{\rm Xe}/N_{\rm nitrogen}$ ratio reaches or exceeds the value determined from solar abundances in a temperature range consistent with the temperature inferred in polar regions and for a plausible amount of liquid on Titan's surface. Thus, a volume of liquid of about $\sim$$6 \times 10^{4}$ km$^{3}$ is required at 90 K to trap enough xenon to explain its apparent atmospheric depletion. The fact that xenon is much more soluble in liquid hydrocarbon than the other noble gases is consistent with their molar masses, i.e. $\sim$39.95 g.mol$^{-1}$, 83.80 g.mol$^{-1}$ and 131.30 g.mol$^{-1}$ for Ar, Kr and Xe, respectively. The most massive element remains the easiest to condense.

\section{Discussion}
\label{sec:disconcl}

  If one considers an atmospheric mole fraction of krypton lower than the detection limit ($y_{\rm Kr} = 10^{-8}$) of the GCMS instrument, the global ratio $N_{\rm Kr}/N_{\rm nitrogen}$ should decrease linearly because $N_{\rm Kr}/N_{\rm nitrogen} \simeq N_{\rm tot, Kr}^{\rm(atm)}/N_{\rm nitrogen}$ (the solubility
of krypton appears to be very low: $N_{\rm tot, Kr}^{\rm(liq)} \ll N_{\rm tot, Kr}^{\rm(atm)}$) and $N_{\rm tot,Kr}^{\rm(atm)} \propto y_{\rm Kr}$ (see Equation~\ref{yatm}).
In contrast, because xenon is much more soluble than the other noble gases in the liquid hydrocarbon, we have $N_{\rm tot, Xe}^{\rm(liq)} \gg N_{\rm tot, Xe}^{\rm(atm)}$. Assuming an atmospheric abundance of xenon lower than the GCMS detection threshold would also lead to a linear decrease of $N_{\rm Xe}/N_{\rm nitrogen}$ because $N_{\rm tot, Xe}^{\rm(liq)} \gg N_{\rm tot, Xe}^{\rm(atm)}$ and $x_{\rm Xe} \propto y_{\rm Xe}$, implying that $N_{\rm tot, Xe}^{\rm(liq)} \propto x_{\rm Xe}$ (see Equation~\ref{Nliq}). We find here a behavior close to Henry's law.\\

The variation of pressure at ground level could also play a role in the dissolution of atmospheric compounds. However, Global Circulation Models (GCM) show that surface pressure fluctuations due to weather conditions or Saturn's tidal effects should not exceed 0.1\% \citep{tokano_neubauer_2002}. On the other hand, using the \textit{Cassini} Synthetic Aperture Radar (SAR), \cite{stiles_etal_2009} have recorded surface height (referenced to the nominal 2575 km radius) in the range -1500 m to +1000 m, yielding {\bf } a maximum altitude difference of $\sim$2500 m. 
We made a test at $T = 90$ K with a pressure 10\% higher than the one measured by \textit{Huygens}, i.e. a value corresponding to the unrealistic case where all the lakes and wet terrains are located at an altitude $\sim $$1500$ m below the \textit{Huygens} landing site. Even with this pressure variation, our calculations show that the fraction of dissolved noble gases is almost unchanged and that the afore-mentioned results remain unaltered.

   We conclude that noble gas trapping in Titan's hydrocarbon lakes and liquid contained in wet surfaces cannot be the unique answer to the problem of their apparent atmospheric depletion. The physical reality is probably composite and several combined effects might play a role with different efficiencies. For example, in order to explain the Ar deficiency, it has been proposed that Titan's building blocks were formed in a relatively warm nebular environment which excluded both argon and molecular nitrogen \citep{owen_1982} or partly devolatilized during their migration within Saturn's subnebula \citep{alibert_mousis_2007,mousis_etal_2009}. If the lakes of Titan are the main sink of atmospheric xenon, then krypton must have remained sequestrated in the interior of Titan because none of the alternative trapping scenarios  cited in the introduction predicts a krypton impoverishment relative to xenon. This study encourages direct measurements by future probes of the noble gas abundances in Saturn's atmosphere and/or in Titan's hydrocarbon lakes.


\begin{acknowledgements}
   We thank an anonymous reviewer for his constructive comments which helped us improve our manuscript. We also thank Bruno B\'ezard and Pascal Rannou for enlightning comments.
\end{acknowledgements}


\begin{thebibliography}{27}
\expandafter\ifx\csname natexlab\endcsname\relax\def\natexlab#1{#1}\fi

\bibitem[{{Alibert} \& {Mousis}(2007)}]{alibert_mousis_2007}
{Alibert}, Y., \& {Mousis}, O. 2007, \aap, 465, 1051

\bibitem[{{Cordier} {et~al.}(2009){Cordier}, {Mousis}, {Lunine}, {Lavvas}, \&
  {Vuitton}}]{cordier_etal_2009}
{Cordier}, D., {Mousis}, O., {Lunine}, J.~I., {Lavvas}, P., \& {Vuitton}, V.
  2009, ApJL, 707, L128, 0911.1860

\bibitem[{{de Kok} {et~al.}(2007){de Kok}, {Irwin}, {Teanby}, {Lellouch},
  {B{\'e}zard}, {Vinatier}, {Nixon}, {Fletcher}, {Howett}, {Calcutt}, {Bowles},
  {Flasar}, \& {Taylor}}]{dekok_etal_2007}
{de Kok}, R. {et~al.} 2007, Icarus, 186, 354

\bibitem[{{Dubouloz} {et~al.}(1989){Dubouloz}, {Raulin}, {Lellouch}, \&
  {Gautier}}]{dubouloz_etal_1989}
{Dubouloz}, N., {Raulin}, F., {Lellouch}, E., \& {Gautier}, D. 1989, Icarus,
  82, 81

\bibitem[{Ferreira \& Lobo(2008)}]{ferreira_lobo_2008}
Ferreira, A., \& Lobo, L. 2008, J. Chem. Thermodyn., 40, 1621

\bibitem[{Ferreira \& Lobo(2009)}]{ferreira_lobo_2009}
------. 2009, The Journal of Chemical Thermodynamics, 41, 809

\bibitem[{{Fulchignoni} {et~al.}(2005){Fulchignoni}, {Ferri}, {Angrilli},
  {Ball}, {Bar-Nun}, {Barucci}, {Bettanini}, {Bianchini}, {Borucki},
  {Colombatti}, {Coradini}, {Coustenis}, {Debei}, {Falkner}, {Fanti},
  {Flamini}, {Gaborit}, {Grard}, {Hamelin}, {Harri}, {Hathi}, {Jernej},
  {Leese}, {Lehto}, {Lion Stoppato}, {L{\'o}pez-Moreno}, {M{\"a}kinen},
  {McDonnell}, {McKay}, {Molina-Cuberos}, {Neubauer}, {Pirronello}, {Rodrigo},
  {Saggin}, {Schwingenschuh}, {Seiff}, {Sim{\~o}es}, {Svedhem}, {Tokano},
  {Towner}, {Trautner}, {Withers}, \& {Zarnecki}}]{fulchignoni_etal_2005}
{Fulchignoni}, M. {et~al.} 2005, Nature, 438, 785

\bibitem[{{Jacovi} \& {Bar-Nun}(2008)}]{jacovi_bar-nun_2008}
{Jacovi}, R., \& {Bar-Nun}, A. 2008, Icarus, 196, 302

\bibitem[{Janssen {et~al.}(2009)Janssen, Lorenz, West, Paganelli, Lopes, Kirk,
  Elachi, Wall, Johnson, Anderson, Boehmer, Callahan, Gim, Hamilton, Kelleher,
  Roth, Stiles, \& Gall}]{janssen_etal_2009}
Janssen, M. {et~al.} 2009, Icarus, 200, 222

\bibitem[{{Lavvas} {et~al.}(2008{\natexlab{a}}){Lavvas}, {Coustenis}, \&
  {Vardavas}}]{lavvas_etal_2008a}
{Lavvas}, P.~P., {Coustenis}, A., \& {Vardavas}, I.~M. 2008{\natexlab{a}},
  \planss, 56, 27

\bibitem[{{Lavvas} {et~al.}(2008{\natexlab{b}}){Lavvas}, {Coustenis}, \&
  {Vardavas}}]{lavvas_etal_2008b}
------. 2008{\natexlab{b}}, \planss, 56, 67

\bibitem[{{Lodders}(2003)}]{lodders_2003}
{Lodders}, K. 2003, \apj, 591, 1220

\bibitem[{{Lorenz} {et~al.}(2008){Lorenz}, {Mitchell}, {Kirk}, {Hayes},
  {Aharonson}, {Zebker}, {Paillou}, {Radebaugh}, {Lunine}, {Janssen}, {Wall},
  {Lopes}, {Stiles}, {Ostro}, {Mitri}, \& {Stofan}}]{lorenz_etal_2008}
{Lorenz}, R.~D. {et~al.} 2008, Geophys. Res. Lett., 35, L02406

\bibitem[{{Mousis} {et~al.}(2009){Mousis}, {Lunine}, {Thomas}, {Pasek},
  {Marb{\oe}uf}, {Alibert}, {Ballenegger}, {Cordier}, {Ellinger}, {Pauzat}, \&
  {Picaud}}]{mousis_etal_2009}
{Mousis}, O. {et~al.} 2009, \apj, 691, 1780, 0810.0308

\bibitem[{{Mousis} \& {Schmitt}(2008)}]{mousis_schmitt_2008}
{Mousis}, O., \& {Schmitt}, B. 2008, \apjl, 677, L67, 0802.1033

\bibitem[{{Niemann} {et~al.}(2005){Niemann}, {Atreya}, {Bauer}, {Carignan},
  {Demick}, {Frost}, {Gautier}, {Haberman}, {Harpold}, {Hunten}, {Israel},
  {Lunine}, {Kasprzak}, {Owen}, {Paulkovich}, {Raulin}, {Raaen}, \&
  {Way}}]{niemann_etal_2005}
{Niemann}, H.~B. {et~al.} 2005, \nat, 438, 779

\bibitem[{{Osegovic} \& {Max}(2005)}]{osegovic_max_2005}
{Osegovic}, J.~P., \& {Max}, M.~D. 2005, J. Geophys. Res-Planet, 110, 8004

\bibitem[{{Owen}(1982)}]{owen_1982}
{Owen}, T. 1982, \planss, 30, 833

\bibitem[{{Owen} \& {Niemann}(2009)}]{owen_niemann_2009}
{Owen}, T., \& {Niemann}, H.~B. 2009, Royal Society of London Philosophical
  Transactions Series A, 367, 607

\bibitem[{{Prausnitz} {et~al.}(1986){Prausnitz}, {Lichtenthaler}, \&
  {Azevedo}}]{prausnitz_1986}
{Prausnitz}, J.~M., {Lichtenthaler}, R.~N., \& {Azevedo}, E.~G. 1986,
  {Molecular Thermodynamics of Fluid-Phase Equilibria}, 2nd edn. (Englewood
  Cliffs: Prentice-Hall)

\bibitem[{{Sloan}(1998)}]{sloan_1998}
{Sloan}, E.~D. 1998, {Clathrates Hydrates of Natural Gases} (New York: Marcel
  Decker)

\bibitem[{{Stiles} {et~al.}(2009){Stiles}, {Hensley}, {Gim}, {Bates}, {Kirk},
  {Hayes}, {Radebaugh}, {Lorenz}, {Mitchell}, {Callahan}, {Zebker}, {Johnson},
  {Wall}, {Lunine}, {Wood}, {Janssen}, {Pelletier}, {West}, {Veeramacheneni},
  \& {Cassini RADAR Team}}]{stiles_etal_2009}
{Stiles}, B.~W. {et~al.} 2009, Icarus, 202, 584

\bibitem[{{Stofan} {et~al.}(2007){Stofan}, {Elachi}, {Lunine}, {Lorenz},
  {Stiles}, {Mitchell}, {Ostro}, {Soderblom}, {Wood}, {Zebker}, {Wall},
  {Janssen}, {Kirk}, {Lopes}, {Paganelli}, {Radebaugh}, {Wye}, {Anderson},
  {Allison}, {Boehmer}, {Callahan}, {Encrenaz}, {Flamini}, {Francescetti},
  {Gim}, {Hamilton}, {Hensley}, {Johnson}, {Kelleher}, {Muhleman}, {Paillou},
  {Picardi}, {Posa}, {Roth}, {Seu}, {Shaffer}, {Vetrella}, \&
  {West}}]{stofan_etal_2007}
{Stofan}, E.~R. {et~al.} 2007, \nat, 445, 61

\bibitem[{{Tegeler} {et~al.}(1999){Tegeler}, {Span}, \&
  {Wagner}}]{tegeler_etal_1999}
{Tegeler}, C., {Span}, S., \& {Wagner}, W. 1999, J. Phys. Chem. Ref. Data, 779,
  829

\bibitem[{{Thomas} {et~al.}(2007){Thomas}, {Mousis}, {Ballenegger}, \&
  {Picaud}}]{thomas_etal_2007}
{Thomas}, C., {Mousis}, O., {Ballenegger}, V., \& {Picaud}, S. 2007, \aap, 474,
  L17, 0708.2158

\bibitem[{{Thomas} {et~al.}(2008){Thomas}, {Picaud}, {Mousis}, \&
  {Ballenegger}}]{thomas_etal_2008}
{Thomas}, C., {Picaud}, S., {Mousis}, O., \& {Ballenegger}, V. 2008, \planss,
  56, 1607, 0803.2884

\bibitem[{{Tokano} \& {Neubauer}(2002)}]{tokano_neubauer_2002}
{Tokano}, T., \& {Neubauer}, F.~M. 2002, Icarus, 158, 499

\end{thebibliography}

\end{document}